# *blockLAW*: Blockchain Technology for Legal Automation and Workflow - Cyber Ethics and Cybersecurity Platforms

**Author(s):**

*Bishwo Prakash Pokharel, Sault College, bishwo889@hotmail.com, https://orcid.org/0000-0001-6515-2105

Naresh Kshetri, Rochester Institute of Technology, naresh.kshetri@rit.edu, https://orcid.org/0000-0002-3282-7331

**Abstract:** In the current legal environment, it is essential to prioritize the protection and reliability of data to promote trust and effectiveness. This study examines how blockchain technology in the form of blockLAW can be applicable to investigate its effects on legal automation, cybersecurity, and ethical concerns. The decentralized ledger and unchangeable characteristics of Blockchain provide opportunities to simplify legal procedures, automate contract execution with smart contracts, and improve transparency in legal transactions. Blockchain is seen as a crucial instrument for updating legal processes while maintaining ethical standards, tackling issues like scalability, regulatory adherence, and ethical dilemmas such as privacy and fairness. The study examines recent developments and evaluates blockchain's impact on legal structures, offering perspectives on its potential to enhance legal procedures and guarantee transparency in legal systems. It further emphasizes blockchain's ability to redefine how legal professionals handle and protect sensitive information, leading to stronger, more effective, and reliable legal procedures. We have also discussed the technological considerations when it comes to blockchain integration into legal systems (like integration planning, implementation strategies, innovations, advancements, trends) with Blockchain Integration Framework for legal systems.

**Keywords:** automation, blockchain, cybersecurity, cyber ethics, legal systems, transparency

## 1. Introduction

Blockchain technology offers an innovative solution for managing and securing data. At its core, it functions as a distributed ledger, recording transactions across numerous computers. Once entered, these transactions become permanent, ensuring that no single party can alter the data retrospectively. This decentralized structure significantly boosts security and trust within the system, as control is not centralized but shared across the network. Essentially, blockchain is a digital platform for recording information, where each event or transaction is logged in a "block". When new transactions occur, the system updates all participants' records. Smart contracts, which employ an unchangeable cryptographic method called hashing, further safeguard data integrity [1].

---

* Correspondence to Bishwo Prakash Pokharel, bishwo889@hotmail.com

Key principles behind blockchain include decentralization, transparency, and immutability. Decentralization disperses authority across a network of nodes, reducing the chance of system failures. Transparency ensures all participants can view transactions, fostering trust. Immutability guarantees that once a transaction is logged, it cannot be changed, ensuring a reliable and trustworthy historical record.

In the realm of legal automation, blockchain holds significant potential. Legal documentation, often plagued by inefficiencies like prolonged processes, high costs, and vulnerability to errors, can benefit from automation tools. Document automation software simplifies tasks like generating contracts, allowing users to input information via questionnaires, and the software instantly produces the necessary documents. Some solutions even create a series of related documents from one initial set of data. These tools either operate as standalone software or integrate with other systems through an API [2].

Legal systems, which often depend on paper documents and manual verifications, are notorious for inefficiencies. Blockchain presents a solution by automating legal processes, enhancing transparency, and securing legal records. Through smart contracts - agreements executed automatically based on pre-coded terms - blockchain reduces the need for intermediaries, speeding up legal proceedings. Furthermore, it improves case management by offering an unamenable record of case histories, documents, and decisions.

This chapter aims to explore how blockchain technology can revolutionize legal workflows. It examines the technical foundations of blockchain, offering insights into its potential applications in the legal industry. The chapter also covers the ethical challenges and cybersecurity concerns that arise when integrating blockchain into legal systems.

## 2. Literature Review

Blockchain technology relies on a combination of technical elements to maintain its decentralized and secure nature. Its core components include blocks, which are structured to hold transaction records. These blocks are connected sequentially through cryptographic hashes, forming a ledger that is resistant to tampering. The use of consensus mechanisms, such as Proof of Work (PoW) or Proof of Stake (PoS), ensures that all participants in the network agree on the blockchain's current state, thus making it more difficult to alter or manipulate.

There are different types of blockchains designed for various applications:
- ***Public blockchains*** are open and fully decentralized, like Bitcoin and Ethereum, and are ideal for situations that require transparency and independence from central authorities.

- ***Private blockchains*** are restricted to certain users, offering greater privacy and control, typically employed in corporate settings where data confidentiality is crucial.
- ***Consortium blockchains*** strike a balance between decentralization and control by being governed by a group of organizations rather than a single entity. These are suitable for collaborations involving multiple stakeholders.

Traditional legal systems face multiple challenges. They are often bogged down by inefficiencies, such as lengthy case processing due to paper-based documentation, manual verification, and multiple intermediaries. This leads to delays, higher costs, and operational bottlenecks. Furthermore, the lack of transparency and accountability in legal procedures fosters errors and potential fraud, diminishing trust in the system. Manual record-keeping is vulnerable to manipulation, making it hard to maintain an accurate, trustworthy record of legal decisions. As legal systems become more digitized, the need for secure and reliable infrastructures grows, with cyber threats posing significant risks to the confidentiality and integrity of legal data.

Blockchain offers various solutions to these legal automation challenges. Smart contracts, which are self-executing contracts with terms encoded directly into code, can help automate legal processes by executing agreements when predefined conditions are met. This reduces the need for intermediaries and speeds up transactions in areas like property sales and compliance monitoring. Additionally, blockchain can enhance case management by providing an immutable ledger of case histories and legal documents, ensuring their integrity and making it easier to verify information throughout legal proceedings. Finally, blockchain can automate compliance checks by embedding regulatory rules within smart contracts, ensuring that transactions meet legal requirements and reducing the risk of non-compliance penalties.

### 3. Ethical Considerations in Blockchain-Enabled Legal Systems

Integrating blockchain into legal systems presents both opportunities and challenges from an ethical standpoint. While blockchain's features can enhance fairness, privacy, and accountability, they also bring complexities that need careful consideration to ensure justice is upheld.

**Privacy concerns:** Blockchain's key feature - data immutability - creates challenges for privacy in legal contexts. Once information is stored, it cannot be altered or removed, which can conflict with privacy rights, such as the "right to be forgotten" in regulations like GDPR. This is particularly problematic when sensitive legal data is involved. Additionally, while transparency in blockchain systems enhances trust, it can compromise confidentiality, making it necessary to incorporate privacy-preserving mechanisms like encryption to protect sensitive information.

**Fairness and bias in automated decision-making:** Smart contracts automate legal processes, but there are concerns over fairness and potential biases. If a smart contract is flawed or biased, it can lead to unjust outcomes without the possibility of human intervention to address special circumstances. Therefore, it's crucial to ensure that algorithms used in legal systems are transparent and accountable, with proper audits and mechanisms to rectify mistakes or biases in automated decisions.

**Equity and accessibility:** Blockchain technology could unintentionally widen the digital divide, making legal services less accessible to marginalized groups who may lack the necessary technological resources or knowledge. To prevent this, legal systems must ensure inclusive access and provide education and training to legal professionals and stakeholders so they can effectively use and oversee blockchain based legal tools.

**Regulatory and compliance issues:** Blockchain operates in a complex regulatory environment with varying laws across different jurisdictions, creating ethical and legal uncertainties. Legal professionals must navigate these challenges to ensure compliance and address cross-border legal issues, which require global frameworks to manage jurisdictional conflicts and the enforcement of legal decisions.

**Enhancing ethical standards with blockchain:** Blockchain can improve ethical standards in legal systems by fostering transparency, accountability, fairness, and privacy. Immutable records enhance accountability by providing a permanent record of actions, reducing the potential for unethical behavior. Transparent processes help to restore public trust in legal systems, while privacy-preserving technologies ensure confidentiality even in open blockchain environments.

To fully harness blockchain's ethical benefits, it's important to implement robust ethical frameworks, particularly around automated decision-making and smart contracts. Additionally, ensuring accessibility through decentralized legal services and promoting education and training will help legal professionals manage ethical challenges more effectively.

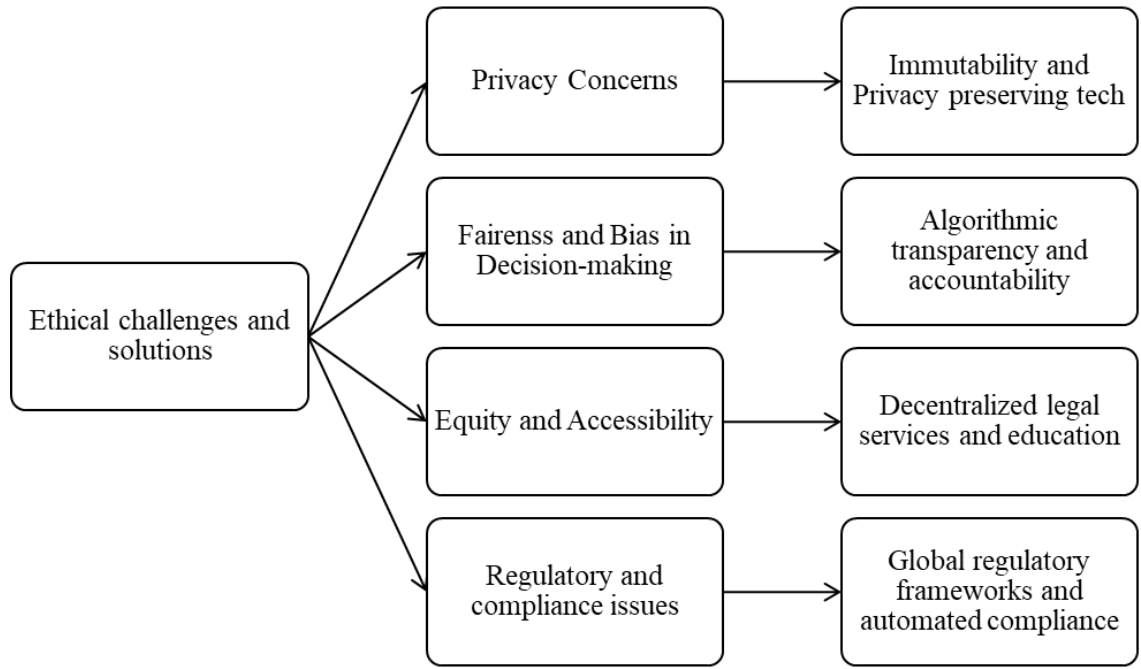

Figure 1: Ethical challenges and solutions in blockchain-enabled legal systems

## 4. Cybersecurity Implications of Blockchain in Legal Systems

Blockchain technology has significant potential to enhance cybersecurity in legal systems, particularly because of its decentralized and immutable nature, making fraud and tampering more difficult. Unlike traditional legal systems, which often rely on centralized databases vulnerable to cyberattacks and manipulation, blockchain disperses data across various network nodes. This distributed structure makes it extremely hard for malicious actors to modify or delete data unnoticed, enhancing the security of legal documents, agreements, and sensitive information [3]. Moreover, cryptographic techniques, such as hashing and digital signatures, bolster the security and reliability of blockchain data, further improving cybersecurity in legal contexts [4].

However, integrating blockchain into legal systems also introduces new cybersecurity challenges. For instance, smart contracts - self-executing agreements coded with specific rules - can streamline legal processes but are susceptible to coding errors and security flaws. If a smart contract is poorly coded, it could result in unauthorized access or fraudulent transactions. Additionally, while blockchain itself is secure, the devices used by legal professionals remain vulnerable to attacks, highlighting the need to secure the entire infrastructure around blockchain [5]. Thus, while blockchain provides strong security advantages, a comprehensive approach is needed to protect both the technology and its supporting systems.

Blockchain serves as a valuable tool for enhancing data security and integrity. Its decentralized ledger system prevents single points of failure, reducing the risk of data breaches or losses common in centralized databases. Each transaction is cryptographically linked to the previous one, making tampering nearly impossible. This ensures that the integrity of the stored data is maintained [5]. Digital signatures further enhance security by ensuring only authorized entities can access or alter data. This robust security framework makes blockchain particularly useful for protecting sensitive legal data, where maintaining data integrity is critical.

By decentralizing control, blockchain reduces potential vulnerabilities and attack surfaces that are often exploited in centralized systems. Traditional systems are susceptible to cyberattacks like ransomware and SQL injection, but blockchain's decentralized nature spreads risk across the network, making such attacks more difficult[6]. Consensus mechanisms like Proof of Work (PoW) and Proof of Stake (PoS) require significant computational resources, further increasing the difficulty of tampering with the ledger. This decentralized design reduces unauthorized access and fraud in legal contexts.

Nevertheless, blockchain is not immune to cybersecurity risks. One potential threat is the 51% attack, where a single entity gains control of the majority of the network's computational power, allowing them to alter transactions. Additionally, attackers can exploit weaknesses in the underlying code of blockchain platforms, leading to breaches [7]. Smart contracts are particularly vulnerable since they execute automatically once deployed, meaning a single coding error can lead to unintended consequences like unauthorized access to funds or data[8]. Other attack methods, like Sybil attacks-where an attacker creates multiple false identities to influence the network - highlight the need for robust security measures in legal systems using blockchain [9].

Mitigating these risks requires ongoing vigilance and best practices. Regular security audits by independent parties help identify and fix vulnerabilities in blockchain platforms [10]. Developers must also follow secure coding standards to avoid introducing flaws in smart contracts and other blockchain applications. Decentralization and redundancy are crucial to prevent single points of failure, and multi-signature transactions, which require approval from multiple parties, add another layer of protection against unauthorized actions [11][12].

The legal landscape for blockchain security is constantly evolving as various jurisdictions adopt different approaches to regulating its use. In Europe, for example, the GDPR has significant implications for blockchain, particularly in terms of data privacy and the right to erasure [13]. In the US, blockchain regulations are more fragmented, with individual states enacting their own laws. However, there is growing recognition of the need for a unified legal framework to address blockchain's unique challenges [14].

As blockchain technology continues to evolve, so must the legal frameworks be governing it. Clear guidelines are needed to ensure smart contracts are legally enforceable and do not lead to unintended consequences. This includes establishing standards for the coding, deployment, and auditing of smart contracts to ensure they are secure and reliable [15]. Additionally, standardizing security practices across the blockchain industry will be crucial to promoting best practices and minimizing vulnerabilities. Regulatory bodies and industry groups should collaborate to develop and disseminate these standards, while legal frameworks must evolve to address decentralized platforms' unique risks, including issues related to jurisdiction, accountability, and consumer protection [16].

## 5. Integration of Blockchain into Legal systems

Blockchain technology holds immense potential to revolutionize legal systems by enhancing transparency, security, and efficiency in legal processes. However, successful integration requires a strategic and careful approach, considering both technological and procedural aspects. This section outlines a framework for the effective implementation of blockchain in legal systems, focusing on critical planning, technological factors, and phased integration strategies.

**Planning for Blockchain Integration**

Before implementing blockchain, it is essential to first assess existing legal workflows. This involves a comprehensive evaluation of current processes, identifying inefficiencies, and determining areas where blockchain can provide the most value. Legal processes are often complex and involve multiple stakeholders, such as lawyers, clients, and regulatory bodies. By understanding these workflows, legal professionals can pinpoint specific challenges, such as delays in document verification, risks of fraud, or issues with data integrity. This assessment also helps in determining whether the legal environment is ready for blockchain implementation and identifies potential obstacles in the transition process [17].

Once an evaluation is complete, clear objectives for blockchain integration must be defined. These goals should align with broader institutional aims, such as improving access to justice, enhancing transparency, or reducing operational costs. For example, blockchain could be used to create tamper-proof legal records, ensuring document integrity, or to automate contract execution using smart contracts, which can significantly reduce the time and resources needed to manage contracts. Defining these objectives provides direction for the integration process and allows for better assessment of its impact and success.

Engaging stakeholders early is critical for successful blockchain adoption. Legal professionals, clients, regulators, and technology vendors all need to be involved in the process to build trust and ensure alignment with their needs. Lawyers, for instance, need to understand how blockchain can improve their practice without compromising ethical responsibilities. Early stakeholder involvement also helps identify resistance to change, gaps in technical knowledge, or other barriers that can be addressed proactively [18].

**Technological Considerations**

Choosing the right blockchain platform is a fundamental decision in the integration process. Different platforms offer varied features, such as consensus mechanisms, scalability, security, and smart contract functionality. Legal systems need to choose a platform that meets their specific requirements. For instance, a private blockchain, which restricts access to authorized participants, may be more suitable for legal applications than a public blockchain [19]. Popular platforms like Ethereum, Hyperledger Fabric, and Corda offer features that can serve legal needs effectively. Key considerations when selecting a platform include its ability to handle large transaction volumes, security features, and compatibility with existing legal technologies [20].

Another important consideration is the integration of blockchain with existing legal systems. Many legal entities rely on legacy systems for case management, document storage, and communication, which may not be readily compatible with blockchain. Creating a seamless integration plan is crucial to ensure that blockchain functions smoothly alongside these systems without causing disruptions. Application Programming Interfaces (APIs) can facilitate the secure exchange of data between blockchain and existing databases. Additionally, it is important to address the legal implications of transferring sensitive data to blockchain systems, ensuring compliance with data protection and privacy laws [21].

Scalability and performance are key to the long-term success of blockchain in legal systems. As the volume of legal transactions and data grows, the blockchain platform must be able to scale to handle increased workloads without compromising speed or efficiency. Scalability has been a challenge for many blockchain platforms, particularly public ones like Bitcoin and Ethereum, where network congestion can lead to delays and higher transaction costs. Solutions such as layer-two scaling (side chains, off-chain transactions) can help distribute workloads and improve performance. As legal systems expand their use of blockchain, they must also ensure the platform maintains robust security and reliability, safeguarding legal processes from cyber threats [22].

**Implementation Strategies**

A phased approach to implementing blockchain is recommended to manage risks and ensure a smooth transition. Incremental adoption allows legal entities to experiment with blockchain in limited use cases before expanding its use more broadly. For instance, blockchain could initially be applied to document verification processes, with wider applications, such as contract management or case tracking introduced as the organization becomes more comfortable with the technology. This gradual approach allows for feedback, identification of potential issues, and necessary adjustments before full-scale implementation. It also improves resource management and minimizes disruption to ongoing legal operations [23].

Education and training are vital for the successful adoption of blockchain. Legal professionals must be familiar with blockchain's principles, benefits, and limitations to effectively incorporate it into their practice. This includes understanding how to use blockchain tools, the role of smart contracts, and the legal implications of blockchain technology. Training programs should be tailored to the needs of different stakeholders, including lawyers, judges, and administrative staff, covering both the technical and legal aspects of blockchain. Continuous learning is crucial as blockchain evolves and new applications emerge [24].

Ensuring regulatory compliance is another critical aspect of blockchain integration. Legal professionals must ensure that blockchain implementations adhere to current regulations, including data protection laws, anti-money laundering (AML) rules, and intellectual property laws. This is particularly important for cross-border transactions and data sharing. Legal entities should collaborate with regulatory authorities to stay updated on legal developments related to blockchain and to ensure compliance. In some cases, new regulations may be necessary to address the unique challenges posed by blockchain and promote its ethical use in legal contexts [25].

**Future Trends and Innovations**

Blockchain's integration into legal systems is likely to evolve alongside other emerging technologies. The convergence of blockchain with artificial intelligence (AI) and machine learning (ML), for instance, offers significant potential. AI and ML can analyze large volumes of legal data, identify patterns, and predict outcomes with high accuracy. When combined with blockchain, these technologies can improve processes like contract analysis or case law research, ensuring data security and reliability. Moreover, blockchain can address AI's "black box" issue by providing transparent and verifiable data used in AI models, fostering greater accountability and trust in AI-driven legal decisions [26][27].

Another important trend is the development of interoperable blockchain networks. The inability of different blockchain platforms to communicate with one another has been a significant barrier to widespread adoption. Interoperability solutions, such as cross-chain communication protocols and blockchain bridges, enable different blockchains to share data and collaborate. This is especially relevant for legal systems that operate across various jurisdictions and need to handle international transactions, ensuring smooth cross-border legal procedures. The recent development of cross-chain communication protocols and blockchain bridges represents significant progress toward achieving this goal. These advancements foster a more interconnected blockchain ecosystem where legal entities from different countries can securely share and verify data, paving the way for a more unified global legal framework [28]. An example of this is the Polkadot network, which supports cross-blockchain interactions, enabling legal firms to streamline international transactions without compromising data integrity or security [29].

Moreover, interoperability is essential for legal systems that wish to embrace blockchain while maintaining flexibility in how they integrate with other technological infrastructures. By ensuring that different blockchain platforms can communicate and share data, legal organizations can expand the use of blockchain in areas such as international arbitration, intellectual property rights, and global compliance, thereby enhancing their operational reach and effectiveness.

**Advancements in Privacy-Preserving Technologies**

With the growing focus on data privacy and security, privacy-preserving technologies play a vital role in the integration of blockchain within legal systems. Innovations such as zero-knowledge proofs (ZKPs), homomorphic encryption, and secure multi-party computation (SMPC) allow blockchain platforms to validate transactions and execute computations without exposing sensitive information. This is crucial in legal contexts where confidentiality is paramount, such as in client-attorney communications or the handling of sensitive legal documents [30].

For instance, Zcash, a digital currency utilizing ZKPs, allows for private transactions that are still verifiable on the blockchain. This principle can be applied in legal systems to manage confidential cases or secure sensitive data while preserving transparency and accountability. Privacy-preserving technologies are expected to become increasingly important as blockchain adoption grows in areas such as intellectual property management, where the need for secure, private communication is critical [31]. These technologies ensure that legal processes can be transparent and accountable while still protecting the rights and privacy of involved parties.

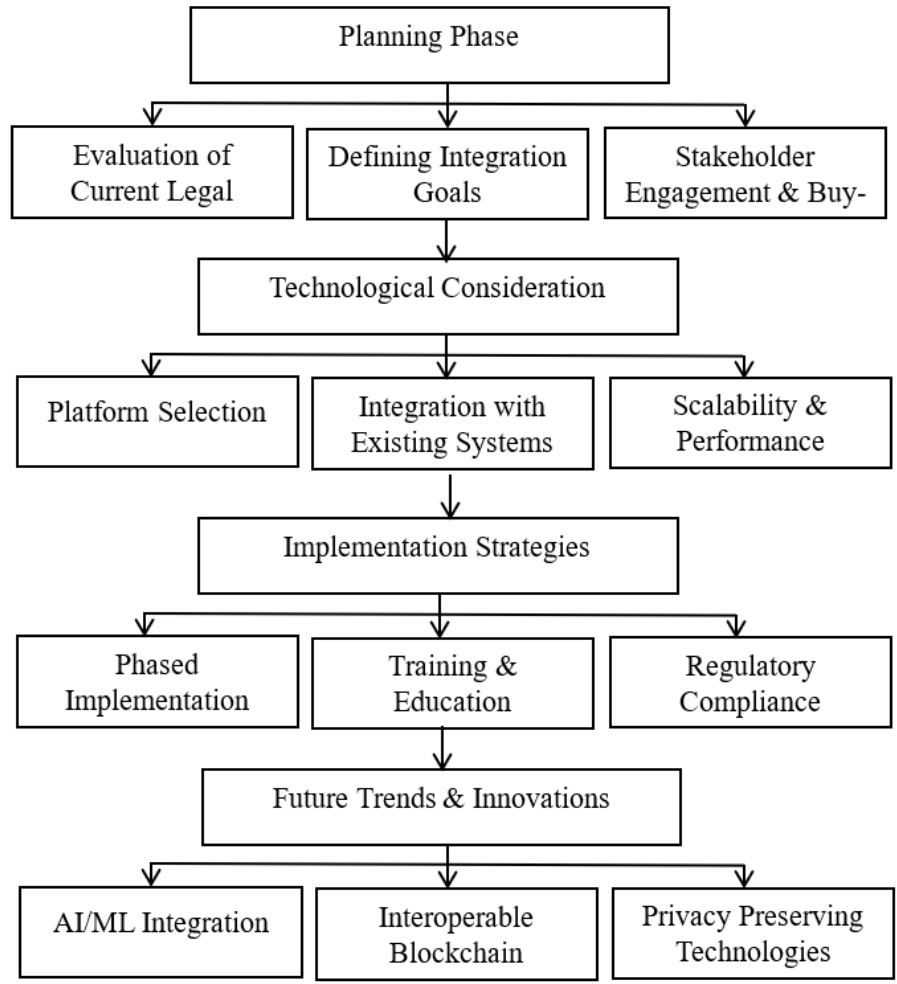

Figure 2: Blockchain Integration Framework for Legal Systems

## 6. Conclusion

Blockchain technology has the potential to drastically change legal systems by improving automation, workflow efficiency, cybersecurity, and ethical standards. By incorporating blockchain technology, legal procedures can enhance transparency, security, and trust by resolving persistent issues like data integrity, unauthorized entry, and ethical responsibility.

Different aspects of blockchain's involvement in legal automation have been examined highlighting the importance of addressing ethical concerns and cybersecurity risks to fully unleash its capabilities. Balancing technological innovation with responsible governance is crucial when addressing the ethical dilemmas associated with integrating blockchain into legal systems, including privacy issues and the requirement for transparency.

Continuous vigilance is necessary to address potential threats such as smart contract weaknesses and Sybil attacks in blockchain-based legal systems, highlighting the ongoing importance of cybersecurity. Nevertheless, by implementing strong mitigation techniques like routine security evaluations, secure programming methodologies, and the utilization of decentralized structures, the legal field can greatly minimize its vulnerability to attacks and improve its overall security.

Moreover, careful planning, stakeholder engagement, and choosing the right technological solutions are essential for the effective incorporation of blockchain into legal systems. The technology is already changing legal practices worldwide with successful use cases like smart contracts in commercial law, blockchain for land registry systems, and decentralized legal case management.

In the future, advancements in combining blockchain with AI and machine learning, creating compatible blockchain networks, and improving privacy-preserving technologies will continue to fuel innovation. These developments will not just confirm blockchain's significance in legal systems but also create opportunities for more secure, efficient, and ethical legal procedures.

In summary, despite the numerous challenges along the path to the widespread implementation of blockchain in legal systems, the potential advantages surpass the risks. By addressing ethical considerations, strengthening cybersecurity measures, and embracing technological advancements, legal professionals can harness the power of blockchain to revolutionize the way legal services are delivered and upheld in the digital age.